\documentclass{ws-ijmpcs}
\usepackage{color}
\usepackage{graphicx}
\usepackage{multirow}
\usepackage{subfigure}
\usepackage{lineno}
\usepackage[rightcaption]{sidecap}

\newcommand{\bfg }{\begin{figure}[htpb]}
\newcommand{\efg }{\end{figure}}
\newcommand{\bmn }{\begin{minipage}}
\newcommand{\emn }{\end{minipage}}
\newcommand{\bt }{\begin{table}[htpb]}
\newcommand{\et }{\end{table}}

\newcommand{\pdA}   {p(d)+A}
\newcommand{\pAu}    {p+Au}
\newcommand{\dAu}    {d+Au}
\newcommand{\pdAu}  {p(d)+Au}
\newcommand{\AuAu}  {Au+Au}

\newcommand{\GeVc}{GeV/$c$ }
\newcommand{\GeVcsq}{GeV/$c^2$ }

\newcommand{ \be }{\begin{equation}}
\newcommand{ \ee }{\end{equation}}
\newcommand{ \bea }{\begin{eqnarray}}
\newcommand{ \eea }{\end{eqnarray}}

\newcommand {\snn}  {\sqrt{s_{_{\rm NN}}}}

\newcommand {\pt}   {p_{T}}

\newcommand {\psiPP}    {\psi_{\rm PP}}
\newcommand {\psiRP}    {\psi_{\rm RP}}

\newcommand {\Bvec} {\vec{B}}
\newcommand {\psiB} {\psi_{B}}

\newcommand {\gSS}  {\gamma_{\rm SS}}
\newcommand {\gOS}  {\gamma_{\rm OS}}
\newcommand {\gdel} {\Delta\gamma}
\newcommand {\dg}   {\Delta\gamma}
\newcommand {\dgscale}  {\dg_{\rm scaled}}

\newcommand {\mult} {N}
\newcommand {\Npv}  {N_{\rm domain}}

\newcommand{\Rmnum}[1]{\expandafter\@slowromancap\romannumeral #1@}

\newcommand {\mean}[1]  {\langle #1\rangle}

\begin{document}

\markboth{Jie Zhao (for the STAR collaboration)}
{Chiral magnetic effect (CME) search at RHIC}

%
\catchline{}{}{}{}{}
%

\title{Chiral magnetic effect search in p(d)+Au, Au+Au collisions at RHIC}

\author{Jie Zhao (for the STAR collaboration)}

\address{Department of Physics and Astronomy, Purdue University, West Lafayette, IN 47906, USA
\\zhao656@purdue.edu}


\maketitle

\begin{history}
\published{Day Month Year}
\end{history}

\begin{abstract}

	The chiral magnetic effect (CME) refers to charge separation along a strong magnetic field of single-handed quarks, 
	caused by interactions with topological gluon fields from QCD vacuum fluctuations. 
	A major background of CME measurements in heavy-ion collisions comes from resonance decays coupled with elliptical flow anisotropy.
	These proceedings present two new studies from STAR to shed further light on the background issue:
	(1) small system p+Au and d+Au collisions where the CME signal is not expected, and 
	(2) pair invariant mass dependence where resonance peaks can be identified.

\keywords{CME; QCD; RHIC.}
\end{abstract}

\maketitle

%
\section{Introduction}
Quark interactions with topological gluon fields can induce chirality imbalance and local parity violation 
in quantum chromodynamics (QCD)\cite{Kharzeev:1998kz,Itoh:1970uw}.  
In relativistic heavy-ion collisions, this can lead to observable electric charge separation along the strong magnetic field, $\Bvec$,
produced by spectator protons\cite{Fukushima:2008xe}.
This is called the chiral magnetic effect (CME).
The commonly used observable to search for the CME-induced charge separation is the three-point correlator difference\cite{Voloshin:2004vk}, $\dg\equiv\gOS-\gSS$ 
; $\gamma=\langle\cos(\phi_{\alpha}+\phi_{\beta}-2\psiRP)\rangle \approx \langle \cos(\phi_{\alpha} + \phi_{\beta} - 2\phi_{c})\rangle/v_{2}$, 
where $\phi_{\alpha}$ and $\phi_{\beta}$ are the azimuthal angles of two charged particles, 
of opposite electric charge sign (OS) or same sign (SS),
and $\psiRP$ 
is that of the 
reaction plane (span by the impact parameter direction and the beam) to which $\Bvec$ is perpendicular on average. 
The latter is often surrogated by the azimuthal angle of a third particle, $\phi_c$, with a resolution correction factor given by the particle's elliptic anisotropy ($v_{2}$).
Significant $\dg$ has indeed been observed in heavy-ion collisions\cite{Kharzeev:2015znc}. 
One of the difficulties in its CME interpretation is a major background contribution 
arising from the coupling of
resonance decay correlations and the $v_{2}$ stemming from the participant geometry\cite{Wang:2009kd,Bzdak:2009fc,Schlichting:2010qia,Adamczyk:2013kcb,Wang:2016iov}.

In non-central heavy-ion collisions, the participant plane azimuthal angle ($\psiPP$), although fluctuating\cite{Alver:2006wh}, is generally 
aligned with the $\psiRP$. 
The $\dg$ measurement is thus $\emph{entangled}$ by the CME and background contributions. 
In small-system \pdA\ collisions, however, the $\psiPP$ is determined purely by geometry fluctuations, 
uncorrelated to the impact parameter or the $\Bvec$ direction\cite{Khachatryan:2016got,Belmont:2016oqp}. 
As a result any CME would average to zero $\dg$ in those collisions. Background sources, on the other hand, 
contribute similarly to small-system \pdA\ and 
heavy-ion collisions  
when measured with respect to the event plane (EP) reconstructed from mid-rapidity particles (as a proxy for the $\psiPP$). 
Due to the fluctuating nature of the small systems \pdA, the EPs reconstructed over a large pseudorapidity ($\eta$) gap can be uncorrelated, 
and therefore measurements with respect to a forward 
EP are sensitive neither to CME nor to the background.
Recent CMS data at the LHC show that the $\dg$ from p+Pb is comparable to that from
Pb+Pb collisions at similar multiplicities\cite{Khachatryan:2016got,Sirunyan:2017quh}, indicating dominant background contribution to the latter. 
In this report, we present results from similar control experiments using p+Au and d+Au collisions at RHIC. 

Because the main backgrounds come from resonance decays, we devise a new analysis approach exploiting the particle 
pair invariant mass, $m_{inv}$, to identify the backgrounds and hence to extract the possible CME signal. 
The $\dg$ signal is reported at large $m_{inv}$ where resonance background contributions are small. 
A two-component model fit to the low mass region is also reported. 

The results reported here use particles in the full TPC 
acceptance ($|\eta| < 1$) for all three particles,
with transverse momentum $0.2< p_{T} <2.0$ \GeVc, and for identified $\pi^{\pm}$ with $0.2< p_{T} <1.8$ \GeVc.
No $\eta$ gap ($\Delta\eta$) is applied between any pair among the three particles for calculating $\mean{\cos(\phi_{\alpha} + \phi_{\beta} - 2\phi_{c})}$. 
The $v_{2}$ used for the resolution correction should, ideally, 
be free of non-flow. We obtain $v_{2}$ by the two-particle cumulant, 
applying a cut of $\Delta\eta > 1.0$ to reduce non-flow contaminations.

\section{Results from small systems}

\begin{figure}[htbp!]
	\centering 
	\includegraphics[width=6.2cm]{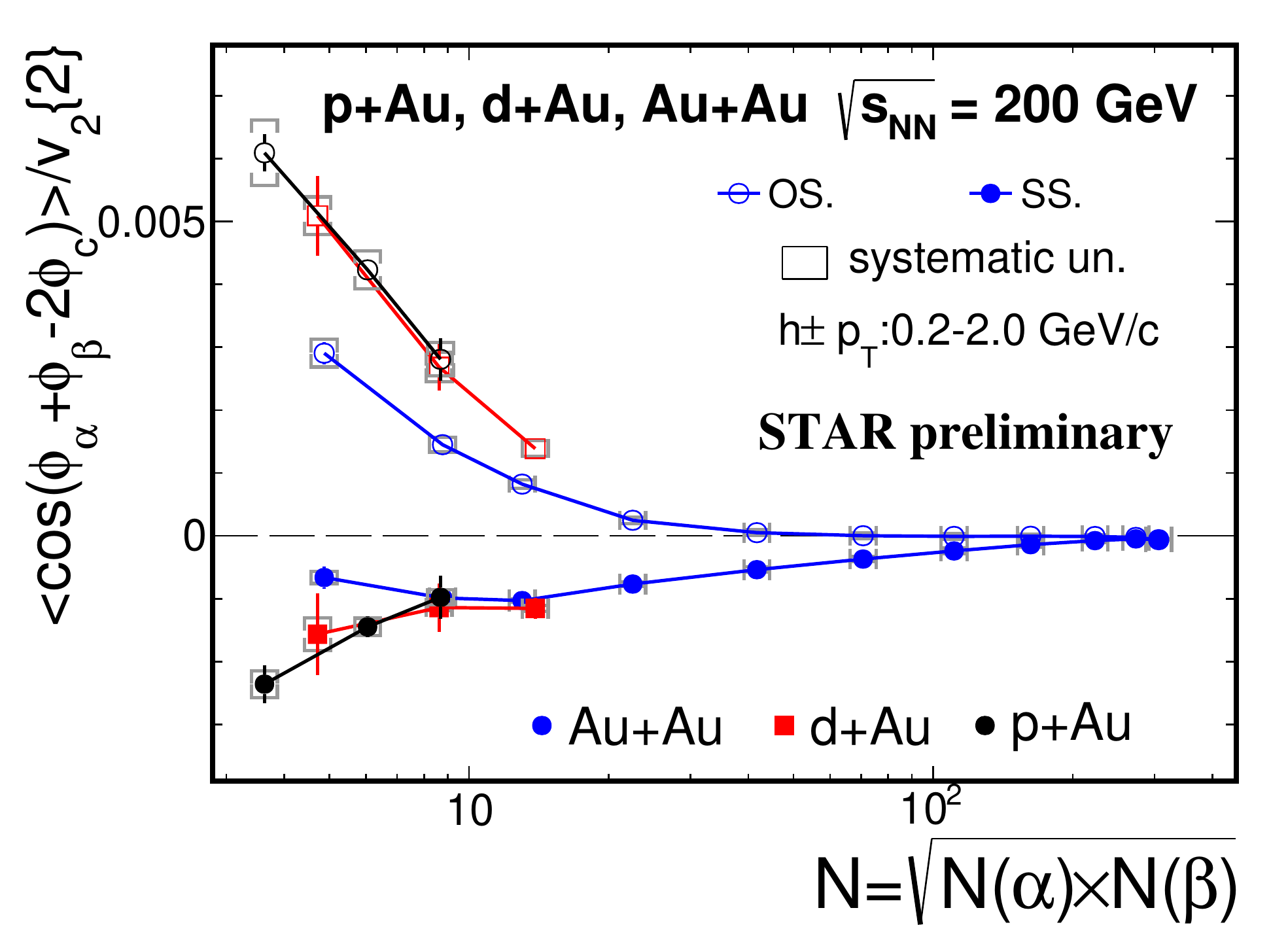} 
	\includegraphics[width=6.2cm]{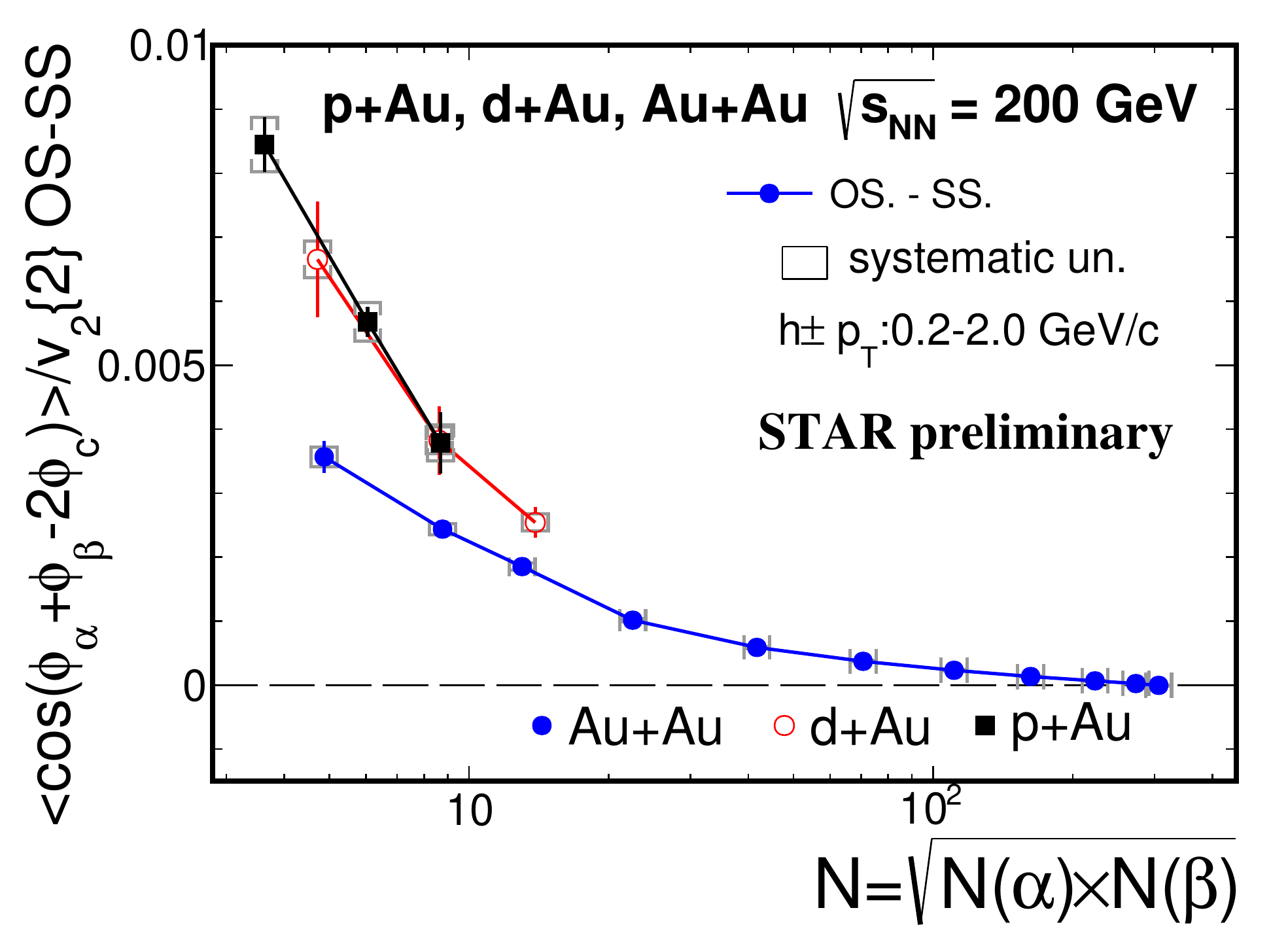}
	\caption{(Color online)
		The $\gSS$, $\gOS$ (Left panel) and $\gdel$ (Right panel) correlators in p+Au and d+Au collisions as a function of multiplicity, compared to those in
		Au+Au collisions. Particles $\alpha$, $\beta$ and $c$ 
		are from the TPC pseudorapidity coverage of $|\eta|<1$ with no $\eta$ gap applied.
		The $v_{2,c}\{2\}$ is obtained by two-particle cumulant with $\eta$ gap of $\Delta\eta > 1.0$. Statistical uncertainties 
		are shown by the vertical bars and systematic uncertainties are shown by the caps.
	}   
	\label{fig1}
\end{figure}

Figure~\ref{fig1}(Left) shows the $\gSS$ and $\gOS$ results as functions of particle multiplicity, $\mult$, in \pAu\ and \dAu\ collisions at $\snn=200$~GeV. 
Here $\mult$ is taken as the geometric mean of the multiplicities of particle $\alpha$ and $\beta$.
The systematic uncertainties on $\gamma$ are estimated by varying the track quality cuts, the correction method used for the detector azimuthal non-uniformity,
and the $\pt$ range of the particle $c$. 
For comparison the corresponding \AuAu\ results are also shown. 
The trends of the correlator magnitudes are similar, decreasing with $\mult$. 
Similar to LHC, the small system \pdAu\ data at RHIC are found to be comparable to Au+Au results at similar multiplicities.

\begin{figure}[htbp!]
	\centering 
	\includegraphics[width=6.2cm]{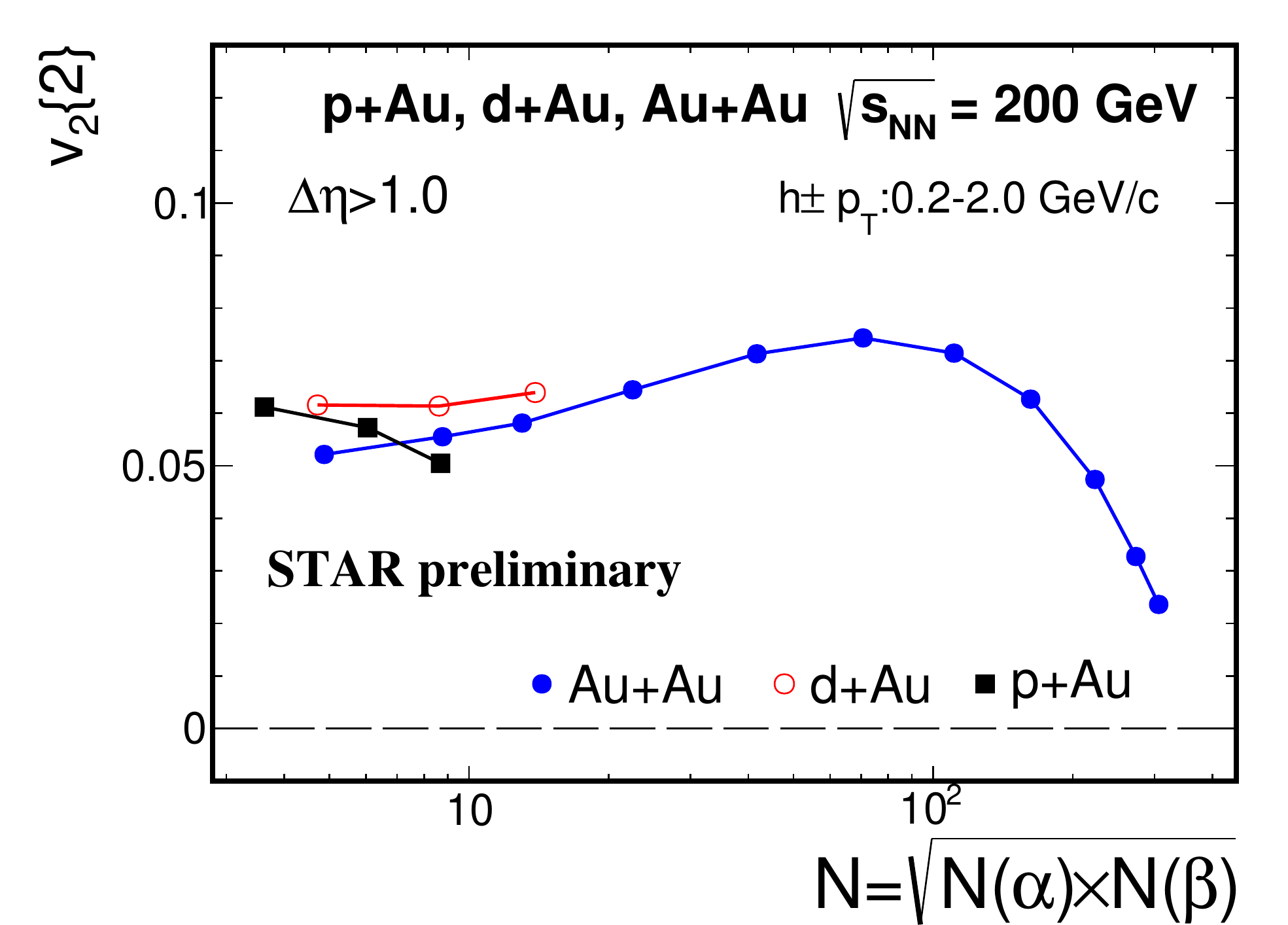} 
	\includegraphics[width=6.2cm]{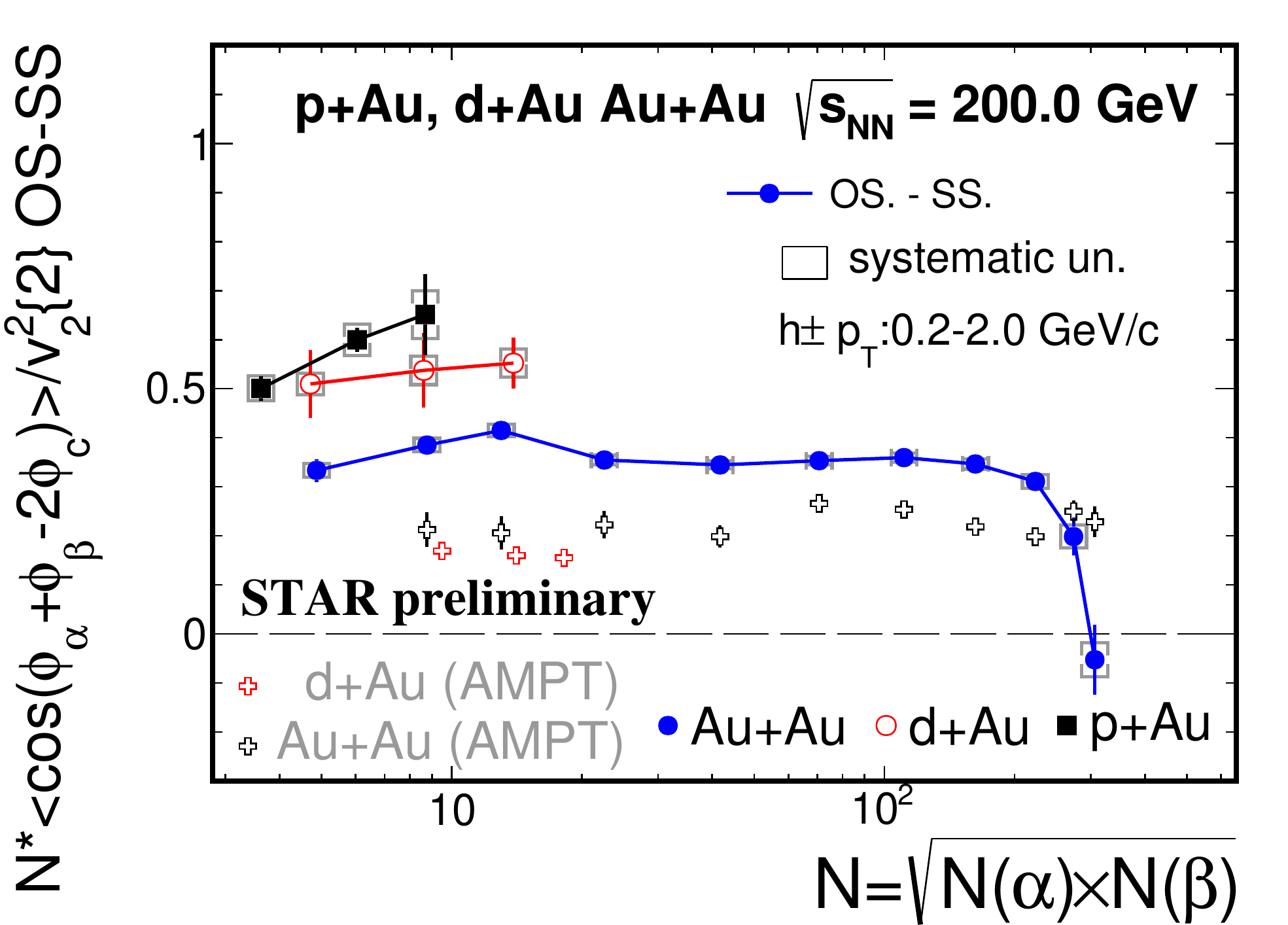}
	\caption{(Color online)
		(Left panel) The measured two-particle cumulant $v_{2}$ of charged particles ($0.2<\pt<2.0$ \GeVc) with $\eta$ gap of 1.0, 
		and (Right panel) the scaled three-point correlator difference in p+Au and d+Au collisions as functions of $\mult$, 
		compared to those in Au+Au collisions. AMPT simulation results for d+Au and Au+Au are also plotted for comparison.
	}   
	\label{fig2}
\end{figure}

Since the \pdAu\ data are all backgrounds, the $\dg$ should be approximately proportional to the $v_2$ of the background sources, 
in turn the $v_2$ of final-state particles. It should also be proportional to the number of background sources, 
and, because $\dg$ is a pair-wise average, inversely proportional to the total number of pairs. 
As the number of background sources likely scales with $\mult$, we have $\dg\propto v_2/\mult$.
Therefore, to gain more insight, 
we study the scaled correlator, $\dgscale=\dg\times\mult/v_{2}\{2\}$.
The $v_{2}\{2\}$ with a $\eta$ gap of 1.0 is shown as a function of $\mult$ in \pAu, \dAu, and \AuAu\ collisions in Fig.~\ref{fig2}(Left).
Figure~\ref{fig2}(Right) shows the $\dgscale$ as a function of $\mult$ in \pdAu\ collisions, and compares to that in \AuAu\ collisions. 
AMPT simulation results for d+Au and Au+Au are also plotted for comparison, which can account for about $2/3$ of the STAR data 
and are approximately constant over $\mult$. 
The $\dgscale$ in \pdAu\ collisions are compatible or even larger than that in \AuAu\ collisions. Since in \pdAu\ collisions only the background is present, the data suggest that the peripheral \AuAu\ measurement may be largely, if not all, background.
For both small-system \pdAu\ and heavy-ion collisions, the $\dgscale$ is approximately constant over $\mult$. 
Given that the background is large suggested by the \pdAu\ data, the approximate $\mult$-independent $\dgscale$ in \AuAu\ collisions is consistent with the background scenario.

In terms of CME, on the other hand, the number of local topological domains, $\Npv$, is likely proportional to the collision volume, so $\Npv\propto\mult$. 
Since the topological charge randomly fluctuates, the charge asymmetry $a_1\propto\sqrt{\Npv}/N$. 
As a result, the $\dg\propto a_1^2$ would be inversely proportional to $\mult$, similar to the background scenario. 
The CME-induced charge separation is expected to be proportional to $\mean{B^2\cos2(\psiB-\psi_2)}$\cite{Kharzeev:2015znc}, 
where $\psiB$ is the azimuthal direction of $\Bvec$. The quantity may have a similar $\mult$-dependence as the $v_2$'s 
but the present theoretical uncertainties are large\cite{Deng:2012pc,Bloczynski:2012en}. 
Thus, the $\mult$-dependence of the CME and the background may, unfortunately, be similar. 
Given the present uncertainties, a small finite CME signal is not disallowed in the measured $\dg$ in heavy-ion collisions.
The present analysis does not currently allow conclusive statements to be made regarding the presence of CME.

\section{Results from the invariant mass method}

\begin{SCfigure}
	\centering 
	\includegraphics[width=6.2cm]{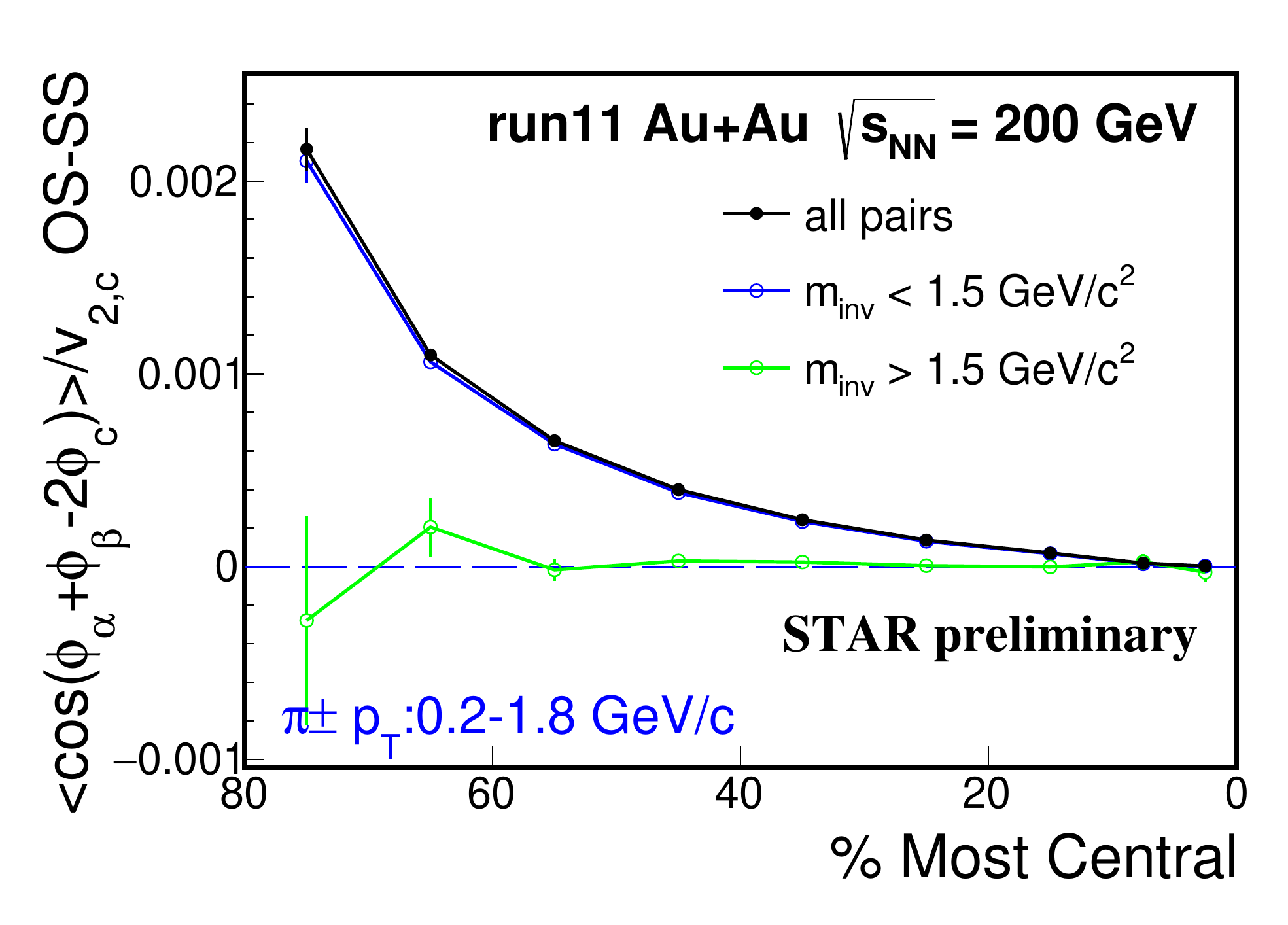}
	\caption{
		The inclusive $\gdel$ over all mass (black) and at $m_{inv} > 1.5$ \GeVcsq (green) as a function of centrality in Au+Au collisions at 200 GeV.  
	}
	\label{fig4}
\end{SCfigure}

As aforementioned, the major background arises from resonance decay correlations coupled with $v_{2}$. 
With increasing mass, the data show that resonance contributions to 
the difference between OS and SS pairs decrease.
It is therefore possible to exclude them ``entirely'' by applying a lower cut on $m_{inv}$. 
Figure~\ref{fig4} 
shows the $\dg$ results with and without applying a cut of $m_{inv}>1.5$~\GeVcsq. 
The results are summarized in Table~\ref{table1}. 
With the $m_{inv}$ cut, the $\dg$ is significantly reduced from the inclusive measurement, 
and is consistent with zero within the current uncertainty.

\begin{table}[ph]
\tbl{The inclusive $\dg$ over all mass and $\dg$ at $m_{inv}>1.5$ \GeVcsq for different centralities in Au+Au collisions at 200 GeV.} 
{\begin{tabular}{lccc}
\hline
\hline
Centrality  &  $\dg$ in all mass (A) &  $\dg$ at $m_{inv}>1.5$ \GeVcsq (B) & B/A \\ 
\hline
50-80\% & $(7.45\pm0.21)\times10^{-4}$   & $(1.3\pm5.7)\times10^{-5}$ & $(1.8\pm7.6)\%$ \\
20-50\% & $(1.82\pm0.03)\times10^{-4}$   & $(7.7\pm9.0)\times10^{-6}$ & $(4.3\pm4.9)\%$ \\
 0-20\% & $(3.70\pm0.67)\times10^{-5}$   & $(-0.1\pm1.8)\times10^{-5}$ & $(-3.8\pm49)\%$ \\
\hline
\hline
\end{tabular}
\label{table1}}
\end{table}


\begin{SCfigure}
	\centering
	\includegraphics[width=6.2cm]{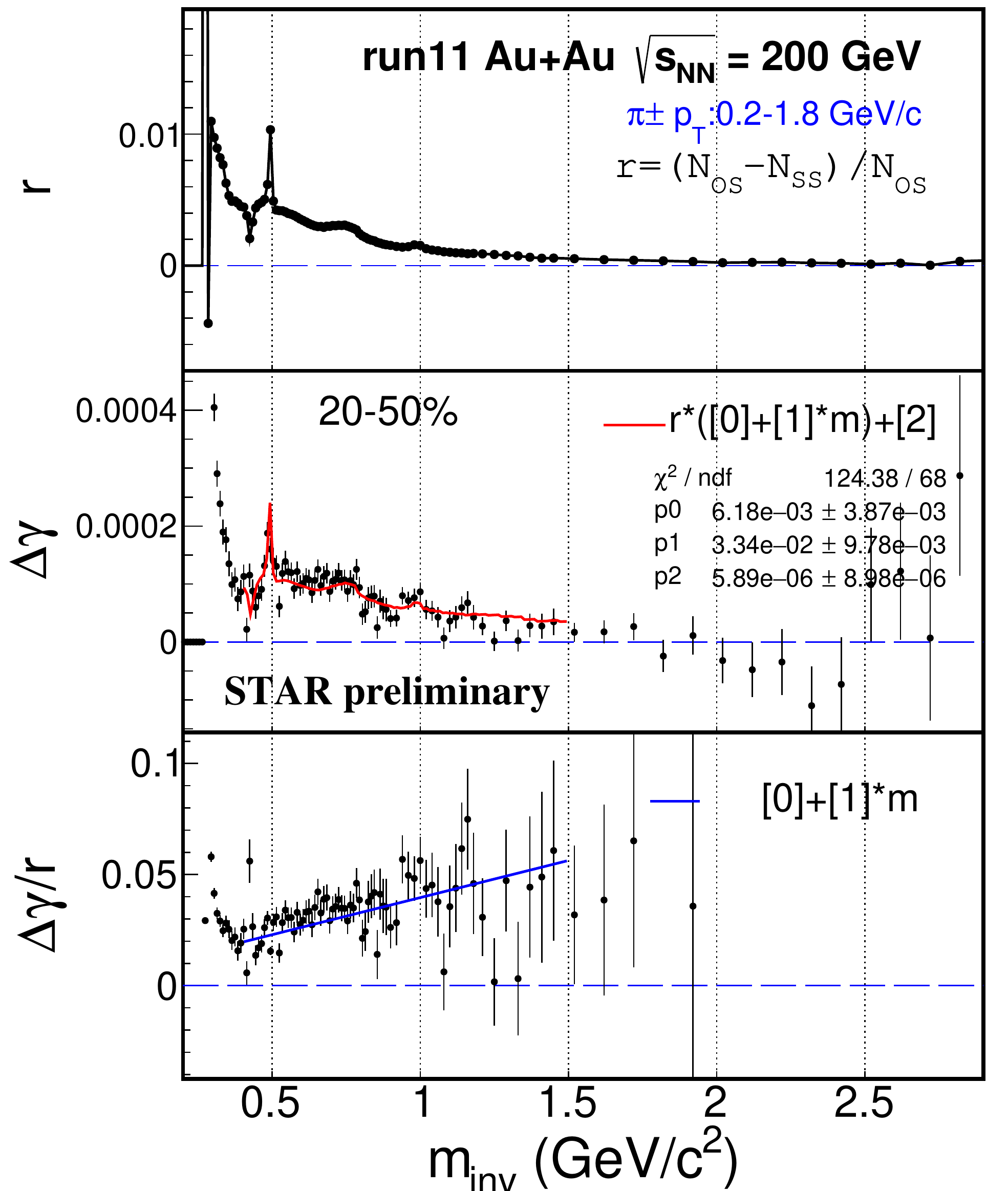} 
	\caption{(Color online)
		Pair invariant mass ($m_{inv}$) dependence of the relative excess of OS over SS charged $\pi$ pairs, $r=(N_{OS}-N_{SS})/N_{OS}$ (top panel), 
		three-point correlator difference, $\gdel=\gOS-\gSS$ (middle panel), and the ratio of $\gdel/r$ (bottom panel) in 20-50$\%$ 
		Au+Au collisions at 200 GeV. Errors shown are statistical.
		The red curve in the middle panel shows the two-component model fit assuming a constant CME contribution independent of $m_{inv}$;
		The blue curve in the bottom panel shows the corresponding resonance response function, $R(m_{inv})$.
	}   
	\label{fig5}
\end{SCfigure}

The CME is expected to be a low $\pt$ phenomenon\cite{Kharzeev:2007jp}; its contribution to high mass may be small. 
In order to extract CME at low mass, resonance contributions need to be subtracted.
To this end, we show in Fig.~\ref{fig5}(a) the relative OS and SS pair abundance difference, $r=(N_{OS}-N_{SS})/N_{OS}$, 
and in Fig.~\ref{fig5}(b) the $\dg$ correlator as a function of $m_{inv}$  
in mid-central (20-50\%) Au+Au collisions at 200~GeV.
The data show resonance structures in $r$ and $\dg$ as functions of $m_{inv}$: 
a clear peak from $K_{s}^{0}$ decay is observed, and possible $\rho$ and $f^{0}$ peaks are visible.
The $\dg$ correlator traces the distributions of those resonances. 

The $\dg$ in Fig.~\ref{fig5}(b) may be composed of two components, a resonance decay background and a CME signal: $\dg=r(m_{inv})\times R(m_{inv}) + \rm{CME}(\it{m_{inv}})$. 
The background depends on $r(m_{inv})$, with a smooth response function $R(m_{inv})$, 
and is therefore peaky in $m_{inv}$. 
We assume that the CME component is smooth in $m_{inv}$. 
If the CME contribution was appreciable, then the ratio of $\dg/r$ shown in Fig.~\ref{fig5}(c) would reveal a structure resembling the inverse shape of $r$\cite{Zhao:2017nfq}. 
However, a more or less smooth dependence is observed;
suggesting insignificant CME signal contributions.

In order to isolate the possible CME from the resonance contributions, 
the two-component model is used to fit the $\dg$ as a function of $m_{inv}$. 
We use the first-order polynomial function for $R(m_{inv})$,
motivated by the data in Fig.~\ref{fig5}(c) and MC simulation\cite{Zhao:2017nfq}.
At present, no theoretical calculation is available on the $m_{inv}$ dependence of the CME, 
therefore we consider a constant CME distribution independent of $m_{inv}$ (Fig.~\ref{fig5}(b)).
Future theoretical calculations of the CME mass dependence would be valuable.


\section{Summary}
The chiral magnetic effect (CME) can produce charge separation perpendicular to the reaction plane. 
Charge separation measurements by the three-point correlator ($\dg$) are contaminated by major backgrounds arising from resonance decay 
correlations coupled with the elliptical anisotropy ($v_2$). 
To further shed light on the background 
and to reduce/eliminate background contamination, 
we have studied the small-system p(d)+Au collisions in comparison to Au+Au collisions, 
and the $\dg$ correlator as a function of the particle pair invariant mass ($m_{inv}$).

With respect to the second-order event plane, the p+Au and d+Au charge dependent 
correlations are backgrounds. Peripheral Au+Au data are similar
to those of p+Au and d+Au. The scaled correlator ($\dg\times\mult/v_2$) from peripheral to mid-central Au+Au
collisions is approximately constant over multiplicity. 
Similar dependence is found in AMPT.
These data indicate a dominant contribution from background to the
peripheral and mid-central heavy-ion collisions, and do not show clear evidence for the
presence of the CME in those collisions.

A new method exploiting $m_{inv}$ is used to identify resonance backgrounds and the possible CME.
In order to exclude the resonance contributions, we apply a lower cut on $m_{inv}$. 
At high mass ($m_{inv}>1.5$~\GeVcsq), $\dg$ is consistent with zero within uncertainty.
In the low mass region ($m_{inv}<1.5$ \GeVcsq), data show resonance structures in $\dg$ as function of $m_{inv}$.
A two-component fit is devised where the background is peaky following the resonance contributions 
and the CME signal is assumed to be smooth in $m_{inv}$. 
A constant distribution for the latter have been studied;
theoretical guidance on its mass dependence would be valuable to further our understanding.


\vspace{3 mm}
$\bold{Acknowledgments}$ This work was partly supported by the U.S. Department of Energy (Grant No. de-sc0012910),
and the NSFC of China under Grant No. 11505073.


\bibliographystyle{woc}
\bibliography{ref}
\end{document}